\documentclass[a4paper]{jpconf}
\usepackage{graphicx}
\begin{document}

\title{Associated Charmonium Production in $p\bar p$ Annihilation}

\author{T.Barnes}

\address{Department of Physics and Astronomy, University of Tennessee, Knoxville TN 37996,\\
Physics Division, Oak Ridge National Laboratory, Oak Ridge TN 37830}

\ead{tbarnes@utk.edu}

\begin{abstract}
In this paper we summarize our recent results for low energy associated charmonium
production cross sections, using 1) crossing symmetry, and 2) an explicit hadronic model.
These predictions are of relevance to the planned charmonium and charmonium hybrid 
production experiment PANDA at GSI.
\end{abstract}

\section{Introduction}
The clear spectrum of states observed in the charmonium system and the straightforward
identification of low-lying states with the predictions of simple $c\bar c$ potential 
models make this an ideal sector for the search for unusual ``extra" states such as 
charmonium hybrids. The PANDA experiment \cite{PandaTechnicalProgress} at GSI plans to
exploit this possibility in a search for charmonia, charmonium hybrids, and other similar 
states in a $p\bar p$ annihilation experiment.

Of course planning for this experiment would be greatly facilitated by experimental data
on and theoretical estimates of these cross sections. Unfortunately, remarkably little is 
known at present. There are a few data points for the single reaction 
$p\bar p \to \pi^0 J/\psi$ from the E760 \cite{Armstrong:1992ae} and 
E835 \cite{Joffe:2004ce,Andreotti:2005vu} experiments at Fermilab, which suggest a scale
of roughly 100~pb near 3.5~GeV, but nothing is known about the cross sections to other 
charmonium states. Here we give results for two estimates of these cross sections, based on
two different theoretical approaches. Where they allow common predictions, we find that 
these methods are reassuringly in agreement to within about a factor of two, and the
explicit hadronic model shows a similar level of agreement with the data. The following 
discussion summarizes these predictions in more detail.

\section{Crossing}
\begin{figure}
\begin{center}
\includegraphics[width=18pc]{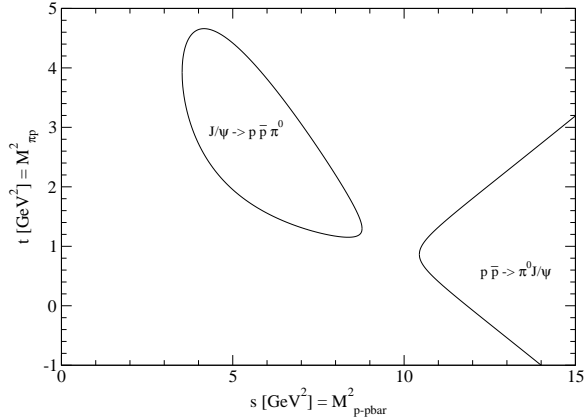}
\caption{Kinematically allowed regimes for the
charmonium production reaction
$p\bar{p}\to \pi^0 J/\psi$
and the related decay
$J/\psi \to \pi^0 p\bar{p}$.}
\label{fig:Dalitz}
\end{center}
\end{figure}

This approach was developed by Lundborg {\it et al.} \cite{Lundborg:2005am}.
The close kinematic proximity of certain charmonium processes with identical 
underlying amplitudes, such as the production process
$p\bar p \to \pi^0 J/\psi$ and the experimentally well studied three body 
decay $J/\psi \to \pi^0 p\bar p$, suggests that knowledge of the decay can be used
to estimate the numerical scale of the production cross section. This is illustrated by 
Fig.\ref{fig:Dalitz}, in which it is evident that an extrapolation from the decay 
Dalitz plot to the production regime may well be feasible.

The differential decay rate and differential production cross section
are the following functions (with an implicit polarization sum) 
of the same invariant amplitude $\mathcal{M}$; 

\begin{equation}
d\Gamma_{\Psi\to m p\bar{p}}
 =\frac{1}{2S_{\Psi}+1}
\frac{1}{(2\pi)^3}\frac{1}{32 M^3} |\mathcal{M} |^2
dm^2_{12}dm^2_{23} \ ,
\label{eq:width}
\end{equation}

\begin{equation}
\frac{d\sigma_{p\bar{p}\to m \Psi}}{dt}=(2S_\Psi+1)(2S_m+1)
\frac{1}{64\pi s}\frac{1}{|p_{1cm}|^2}|\mathcal{M}|^2,
\end{equation}
Here $m$ is a generic meson with spin $S_m$ and $\Psi$ is a generic charmonium (or related)
state. Ideally one can develop a detailed model
of $\mathcal{M}$ from a study of the decay Dalitz plot, which can then be extrapolated 
to predict the charmonium production cross section. However
a much simpler qualitative cross section estimate can be written if the Dalitz plot 
does not show much structure, which is the constant amplitude approximation. If we assume
that $|\mathcal{M}|^2$ is constant, it can be eliminated between the decay and 
cross section formulas, which gives a very simple cross section estimate in terms of the 
three-body decay rate:
\begin{equation}
\sigma_{p\bar{p}\to m \Psi}=
\frac{12\pi^2M^3}{A_D}\Gamma_{\Psi\to m p\bar{p}}
\frac{p_{3cm}}{p_{1cm}s}
\label{eq:connect}
\end{equation}
where $A_D$ is the area of the decay Dalitz plot.

\begin{figure}[h]
\begin{center}
\includegraphics[width=18pc]{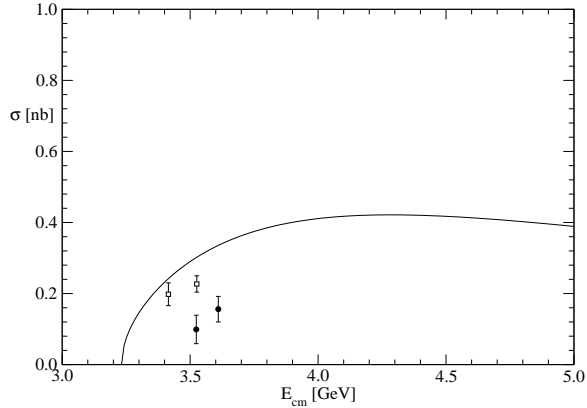}
\caption{Constant-amplitude prediction for
$\sigma(p\bar{p}\to  \pi^0 J/\psi)$.
The data points are Fermilab measurements from
E760 \cite{Armstrong:1992ae} (filled)
and E835 \cite{Joffe:2004ce} (open).
There are additional experimental results for this reaction from E835 
\cite{Andreotti:2005vu}
which have not yet been published as a physical cross section.}
\label{fig:crosssection}
\end{center}
\end{figure}

This simple constant-amplitude prediction for the cross section for
$p\bar{p}\to \pi^0 J/\psi$ is shown in Fig.\ref{fig:crosssection}. 
This is the single experimentally measured case; as we noted previously,
there are a few data points for this cross section from the E760 
\cite{Armstrong:1992ae} and E835 \cite{Joffe:2004ce,Andreotti:2005vu} 
experiments at Fermilab, which are also shown in the figure. Evidently there
is indeed rough agreement between theory and experiment, which suggests that
extrapolation from data on generic decays of the form 
$\Psi \to m p \bar p$ to the corresponding associated production cross section
$p \bar p \to m \Psi $ is a useful approach. 

The corresponding results for a range of $J/\psi$ and $\psi'$ decays are given 
in Ref.\cite{Lundborg:2005am}. One concern in applying this approach is that 
some decay processes may be dominated by strong $N^*$ resonance bands, in which case 
extrapolation to the production cross section would be inaccurate. An example 
is $J/\psi \to \eta p \bar p$, which presumably has large contributions from
the $N^*(1535)$. Extrapolation of this and other decay 
processes to production cross section should generally be accompanied by 
careful studies of the decay Dalitz plot which allow $N^*$ resonance contributions.

\section{Hadronic Model}
\begin{figure}[h]
\begin{center}
\includegraphics[width=18pc]{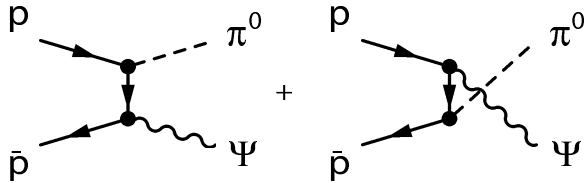}
\end{center}
\caption{Feynman diagrams assumed in this model of
the generic reaction $p\bar p \to \pi^0 \Psi$.}
\label{fig:diags}
\end{figure}

An attractively simple dynamical model which we have recently developed 
\cite{Barnes:2006ck} assumes that the associated
production of a charmonium or charmonium hybrid state $\Psi$ with a $\pi^0$ 
in $p\bar p$ annihilation takes place 
through pion emission from an incoming $p$ or $\bar p$ line. 
The process $p\bar p \to \pi^0 \Psi$ 
at leading order then involves the two Feynman diagrams of Fig.\ref{fig:diags}, in which
the only {\it a priori} unknown quantity is the $\Psi - p\bar p$ vertex.

The use of a similar model, with a PCAC axial vector coupling for the $pp\pi^0$ vertex,
was earlier suggested by Gaillard, Maiani and Petronzio \cite{Gaillard:1982zm}. 

We have used this model (with a simpler pseudoscalar $pp\pi^0$ vertex) to calculate
the cross sections for $p\bar p \to \pi^0 \Psi$ for a much wider range of light 
charmonia, $\Psi = \eta_c, J/\psi, \chi_0, \chi_1$ and $\psi'$. Numerical results for these
cross sections require the strength of the coupling $\Psi - p\bar p$, which we take from 
the measured annihilation widths to $p\bar p$. (We do not consider the $\chi_2$ because 
of the more complicated vertex, and other charmonium states do not have known $p\bar p$ 
couplings.) Analytical results for all these total and differential cross sections 
and the numerical results shown here are given in Ref.\cite{Barnes:2006ck}.

\begin{figure}[h]
\begin{center}
\vskip 0.5cm
\includegraphics[width=18pc]{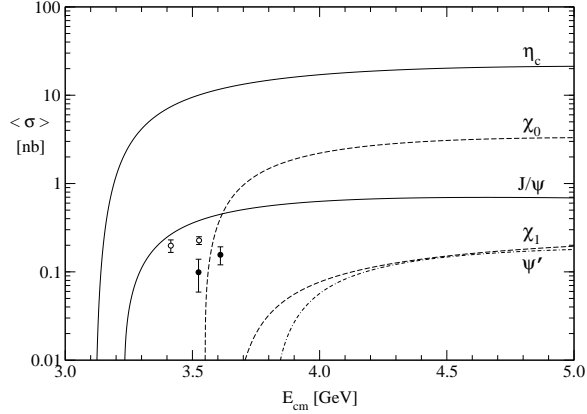}
\end{center}
\caption{\label{fig:csecs}Predicted unpolarized total cross sections for the processes
$p \bar p \to \pi^0 \Psi$, where $\Psi = \eta_c, J/\psi, \chi_0,
\chi_1$ and $\psi'$. The data points are the E760 (filled) and E835 (open) 
experimental results for $\Psi = J/\psi$.}
\end{figure}

Our results for these cross sections are shown in Fig.\ref{fig:csecs}. First, note that
the single measured case of  $p\bar p \to \pi^0 J/\psi$ is indeed in rough (factor of two)
agreement with the predictions of this dynamical model. This suggests that the simple 
Feynman diagrams of Fig.\ref{fig:diags} may well give a reasonably accurate 
description of these 
processes, which is very encouraging. Application of the model to other charmonium and 
charmonium hybrid states only requires knowledge of the coupling $\Psi - p\bar p$.

\begin{figure}
\begin{center}
\vskip 0.6cm
\includegraphics[width=18pc]{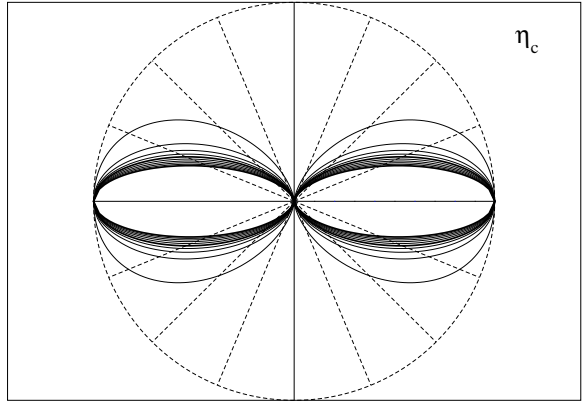}
\end{center}
\caption{\label{fig:W_etac}The corresponding angular distribution for the reaction 
$p \bar p \to \pi^0 \eta_c$, for E$_{cm} = 3.2-5.0$~GeV in steps of 0.2~GeV.}
\end{figure}
The most remarkable predictions of the model are the very large 
cross sections for the associated production of the $\eta_c$ and $\chi_0$ states
relative to the $J/\psi$ (see Fig.\ref{fig:csecs}). In particular, the cross section for 
$p\bar p \to \pi^0 \eta_c$
is predicted to be about a factor of 30 larger than the 
$p\bar p \to \pi^0 J/\psi$ cross section in the range of $\sqrt{s}$ relevant to PANDA.
The reason for the large $\eta_c$ and $\chi_0$ couplings to $p\bar p$ is not known,
but it may be that $p\bar p$ annihilation through $gg$ final states is significantly
larger than the $ggg$ states that couple to the $J/\psi$. This suggests that production 
of some hybrid states may be enhanced at PANDA over $c\bar c$, since $gg$ can couple to
the hybrid basis state $c\bar c g$ more easily than to $c\bar c$ itself \cite{Swanson}.

This model also gives interesting predictions for the angular distributions of charmonia
produced in $p\bar p$ associated with a $\pi^0$. As an example, the c.m. frame 
angular distribution for the largest cross section, $p\bar p \to \pi^0 \eta_c$,
is shown in Fig.\ref{fig:W_etac}; this angular distribution 
has a very characteristic node at $\theta = \pi/2$, due to a $t\leftrightarrow u$ 
antisymmetry in this model. Since the angular distributions we find for
$p\bar p \to \pi^0 \Psi$ are sensitive to the quantum numbers of the charmonium
state $\Psi$, they could be used as tests of the accuracy of this simple hadronic model.

\section{Acknowledgments}

I would like to thank the organizers for the opportunity to present this 
contribution at GHP2006.
This research was supported in part by
the U.S. National Science Foundation through grant NSF-PHY-0244786 at the
University of Tennessee, and the U.S. Department of Energy under contract
DE-AC05-00OR22725 at Oak Ridge National Laboratory.

\end{document}